\documentclass[fleqn,twoside]{article}
\usepackage{espcrc2}

\usepackage{graphicx}
\usepackage[figuresright]{rotating}

\def\gsim{\,\,\rlap{\raise 3pt\hbox{$>$}}{\lower 3pt\hbox{$\sim$}}\,\,}
\def\lsim{\,\,\rlap{\raise 3pt\hbox{$<$}}{\lower 3pt\hbox{$\sim$}}\,\,}

\newcommand{\vev}[1]{\langle {#1} \rangle}
\newcommand{\ord}[1]{\mathcal{O}{(#1)}}
\newcommand{\beq}{\begin{equation}}
\newcommand{\eeq}{\end{equation}}
\newcommand{\fb}[1]{{#1}~fb$^{-1}$}

\newcommand{\AmS}{{\protect\the\textfont2
  A\kern-.1667em\lower.5ex\hbox{M}\kern-.125emS}}

\title{Echoes from a Warped Dimension}

\author{Hooman Davoudiasl\address{Department of Physics, Brookhaven National Laboratory\\
Upton, NY 11973, USA}\thanks{Work supported by the US Department of
Energy under Grant Contract DE-AC02-98CH10886.}}
       
\begin{document}

\begin{abstract}
The Randall-Sundrum (RS) model, based on a slice of warped 5D spacetime, 
was originally introduced to explain the apparent hierarchy between the scales of weak 
and gravitational interactions.  Over the past decade, this model has been 
extended to provide a predictive theory of flavor, as well as to address various 
constraints from precision data.  In this talk, we will present a brief review of 
the RS model and some of its extensions.  A survey of some key signatures of 
realistic warped models and the experimental challenges they pose will be presented. 
We will also consider truncated warped scenarios that address smaller hierarchies 
and discuss why their discovery can have improved prospects, at the LHC.
\vspace{1pc}
\end{abstract}

\maketitle

\section{Introduction}

The following is based on an invited talk given by  the author at 
The International Workshop on 
``Beyond the Standard Model Physics and LHC Signatures (BSM-LHC)," 
held at Northeastern University, Boston, MA, USA, June 2-4, 2009.

In the Standard Model (SM), electroweak symmetry breaking (EWSB) 
is achieved by the scalar Higgs doublet condensate $\vev{H}=v$.   
Given the quadratic sensitivity of Higgs mass to quantum corrections from 
an arbitrarily high mass scale $M$, one is faced with the question of why 
$v \ll M$.  This is the general nature of what is referred to as the hierarchy 
problem.  In particular, if $M$ is near the Planck scale 
$M_P\sim 10^{19}$~GeV, one is compelled to explain why $v/M_P \sim 10^{-16}$.  
Over the years many ideas have been advanced to explain the hierarchy, including 
strong dynamics, supersymmetry, and  more recently extra dimensions.  In this 
presentation, we will focus on a proposal for bridging the energy interval between 
a $\mu$~g and 1 erg, through a warped extra dimension.  

\section{The Randall-Sundrum Model}

Warped 5D models that we will examine here are generally based on 
the Randall-Sundrum (RS) model \cite{Randall:1999ee}, which is a slice of the AdS$_5$ spacetime 
bounded by two 4D Minkowski walls at $y=0, \pi r_c$, the UV and IR branes, respectively, 
along the fifth dimension.  The RS metric is given by:
\beq
ds^2 = e^{-2 k y} \eta_{\mu \nu} \,dx^\mu dx^\nu - dy^2, 
\label{metric}
\eeq 
where $k$ is the curvature scale and is typically assumed to be somewhat smaller 
than the 5D fundamental scale $M_5\sim M_P$.  Physical mass scales get redshifted in 
this background as one goes from the UV brane to the IR  brane.  In particular, 
if the 5D Higgs condensate $v_5 \sim k$ is IR-localized, the observed 
4D value will be obtained from $e^{-k r_c \pi} \vev{H_5} $ with $k r_c \pi \approx 36$, 
hence resolving the hierarchy problem.  We note that based on the AdS/CFT 
correspondence \cite{Maldacena:1997re}, there is a connection between 
certain 4D strong dynamical models and RS-type
models \cite{RSCFT}, but we will not examine this subject any further here.

The most distinct signature of the original RS model is the appearance of a tower 
of spin-2 resonances, corresponding to the Kaluza-Klein (KK) states of the 
5D graviton.  These KK gravitons have masses and couplings governed by the 
TeV scale, and would couple to all SM fields universally.  This gives rise to striking  
predictions for collider experiments \cite{Davoudiasl:1999jd}.  Another aspect of the original 
RS phenomenology is related to the required stabilization of the size 
of the fifth dimension \cite{Goldberger:1999uk}, whose quantum fluctuations gives rise to 
the {\it radion} scalar, with properties 
very similar to the SM Higgs \cite{Goldberger:1999un}.  The radion 
can mix with the Higgs, through a possible scalar-curvature 
coupling \cite{Csaki:2000zn}.  

\begin{figure}[htb]
\includegraphics[width=8cm]{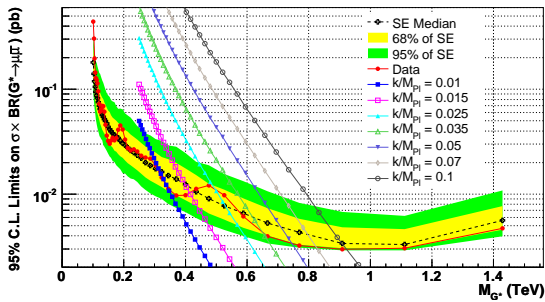}
\caption{
CDF limits, at $95\%$ confidence level, 
on the product of cross section and di-muon branching ratio,  
for $G^1$ in the original RS model.   Theoretical cross sections, as well as  
the expected limits from simulated experiments (SE) are shown; from Ref.~\cite{Aaltonen:2008ah}}
\label{CDF}
\end{figure}

\begin{figure}[htb]
\includegraphics[angle=-90,width=7.4cm]{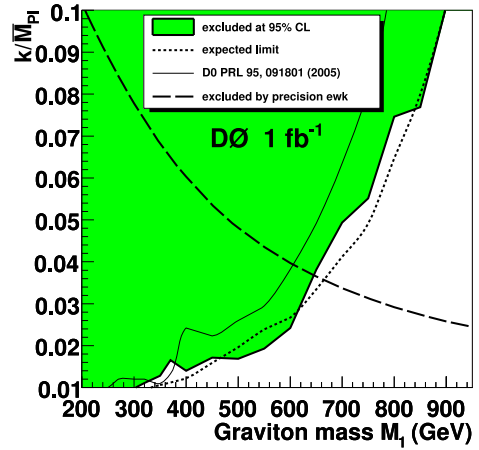}
\caption{D0 bounds on $G^1$, using the $e^+e^-$ and $\gamma \gamma$ 
channels, from Ref.~\cite{Abazov:2007ra}.}
\label{D0}
\end{figure}

The Tevatron experiments have searched for the first graviton KK mode $G^1$ in the original model and 
have not found a signal, resulting in the bounds presented in Figs.\ref{CDF} and \ref{D0}, from the 
CDF (\fb{2.3}) \cite{Aaltonen:2008ah} and D0 (\fb{1}) \cite{Abazov:2007ra} 
experiments, respectively.  Roughly speaking, the current data disfavors 
a $G^1$ lighter than 900~GeV, for $k/M_P = 0.1$, at 95\% confidence level.

The LHC experiments ATLAS and CMS have examined their reach for $G^1$ in the original 
RS model.  With \fb{100} and $k/M_P = 0.1$, the ATLAS experiment 
\cite{Allanach:2002gn} expects to be able to 
discover $G^1$, in the $e^+ e^-$ channel, up to a mass of 3.5~TeV, 
whereas the CMS reach is somewhat better (about 4~TeV), 
in the di-muon channel \cite{CMS_RS}.

\section{Warped Models of Hierarchy and Flavor} 

Although the original RS model provides an interesting resolution of the 
hierarchy problem and had distinct testable predictions, it also gives rise to some 
phenomenological problems.  For one thing, the model has a (quantum gravity) cutoff scale 
of $\sim 1$~TeV, which is quite insufficient to suppress higher-dimensional operators 
that would cause deviations from precision EW or flavor data.  Raising the cutoff scale, 
which is close to the KK scale of the IR brane, will introduce hierarchies that the model 
was meant to address.  Also, the SM flavor structure still remains a mystery in this setup; 
while this is a short-coming shared by many models of hierarchy, it would be 
desirable to use the 5D bulk to address the flavor problem as well (as was done in the 
context of flat backgrounds \cite{ArkaniHamed:1999dc}).

\begin{figure}[htb]
\includegraphics[width=7.4cm]{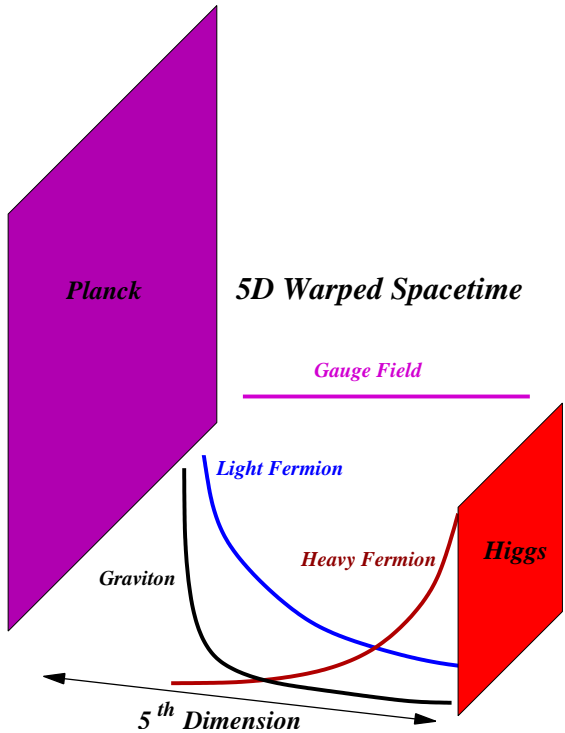}
\caption{Schematic representation of a 5D warped model of hierarchy and flavor.}
\label{RS5DSM}
\end{figure}

In fact, to address the hierarchy problem, it is sufficient to keep 
the Higgs near the IR brane and the rest of the SM can be allowed to 
reside in all five dimensions \cite{Goldberger:1999wh}.  
The gauge fields \cite{Davoudiasl:1999tf,Pomarol:1999ad} 
and fermions \cite{Grossman:1999ra} can be moved to 
the bulk, where 5D masses for the latter can give rise to a natural 
geometric mechanism, based on the overlap of 
5D profiles with the IR-localized Higgs, to generate 4D fermion masses over a 
wide range of values.  Here, the light fermions are localized near the UV brane, 
while heavy fermions are IR-localized.  A schematic representation of this setup 
is provided in Fig.~\ref{RS5DSM}.   
In addition to generating small masses, UV-localization of the light fermions 
also raises the effective cutoff scale for operators composed of these fields  far 
above the TeV regime \cite{Gherghetta:2000qt}.  
This feature provides for an efficient means of suppressing 
unwanted operators, such as those mediating flavor changing neutral 
current (FCNC) processes, related to the tightly constrained light flavors.  Hence, 
5D warped scenarios can also provide natural and predictive models of flavor, whose 
typical signals are expected to arise near the weak scale.

This simultaneous resolution of the hierarchy and flavor puzzles, within 
the extended warped models, comes at a price.  That is, in these models, 
the new states become more elusive in collider experiments, and in particular, 
at the LHC.  If we denote a generic SM gauge coupling by 
$g_{SM}$, the new couplings are roughly given as follows 
\vskip 0.1cm

$\bullet$ Gauge KK couplings:
\vskip0.2cm

\begin{flushleft}
- UV-brane ({\it e.g.} $e$, $u$): $\sim g/\sqrt{k \pi r_c}$

- IR-brane ({\it e.g.} $H$, $t_R$): $\sim g \sqrt{k \pi r_c}$

\end{flushleft}

\vskip0.5cm

$\bullet$  Graviton KK couplings in $\sim$TeV$^{-1}$ units:
\vskip0.2cm

\begin{flushleft}

- Light fermions: $\sim$ Yukawa (overlap with the IR-localized graviton KK mode)

- IR-brane ({\it e.g.} $H$, $t_R$): $\sim 1$

- Gauge fields ($g$, $\gamma$): $\sim 1/(k\pi r_c)$ (volume suppressed).

\end{flushleft}

We see that KK couplings to light SM fields, 
important for production ({\it e.g.} light quarks) 
and clean detection ({\it e.g.} the photon) at colliders, are suppressed in 
models that yield realistic 4D flavor patterns.  Instead of the usual golden channels 
that are easy to detect, the KK modes most strongly couple to heavy SM fields, such 
as top quarks and the Higgs sector, including the longitudinal gauge fields 
$W^\pm_L$ and $Z_L$, which require more complicated event reconstruction.  In particular, 
the projections regarding the reach for signals of warped scenarios must be revised 
for models of hierarchy and flavor.  However, before discussing the experimental 
prospects for the simplest warped models that also accommodate flavor, 
let us briefly review some of the relevant constraints from 
precision data and their implication for collider searches.

Even though fermion localization alleviates some of the problems with warped models a great deal, 
there are still some rather stringent experimental bounds that could push the masses of KK modes 
to scales high enough to render RS-type models unnatural from the point of view of the 
hierarchy and, in any case, inaccessible at the LHC.  Of these, the constraints from oblique corrections 
can be brought under control by enlarging the bulk gauge group 
to $G_c = SU(2)_L\times SU(2)_R\times U(1)_X$, for gauge KK 
masses $m_{KK} \gsim 3$~TeV \cite{Agashe:2003zs}, due to the custodial 
protection provided by $G_c$.  The gauge KK masses would be pushed 
to about 5~TeV or more, if we also include the constraints from the measured $Z b {\bar b}$ 
coupling, unless we enhance the symmetry to 
$G_c \times Z_2$ \cite{Agashe:2006at}.  In addition, important 
bounds come from gluon KK exchange processes \cite{Agashe:2004cp} contributing to 
the $\epsilon_K$ parameter of $K-{\bar K}$ mixing, the most severe of which comes from 
$(V-A)\times (V+A)$ operators \cite{Beall:1981ze,Bona:2007vi}, 
that would require approximately $m_{KK} \gsim 20$~TeV \cite{Csaki:2008zd}.  Without 
further tuning of parameters or extra flavor structure (see, for example, Ref.~\cite{Fitzpatrick:2008zza})  
these bounds would not allow for observable KK states at the LHC.  Nonetheless, in the following 
discussion we will assume that remedies are in place to address the flavor bounds and we will 
examine the discovery prospects in relation to constraints that are addressed by imposing $G_c$ 
in the bulk.

\section{Realistic Bulk Flavor and LHC Signals}

Next, we will give a brief summary of some of the works  that have examined the 
discovery reach for 5D warped models with localized fermions that can explain 
the 4D SM flavor structure (for a more detailed survey of warped collider 
phenomenology and additional references, see, for example,  
Ref.~\cite{Davoudiasl:2009cd}).  These results 
typically apply to the simplest models, endowed with an IR-localized  
Higgs fields, and only consider the SM decay modes of the KK  states.  
However, the width of these states can receive important contributions 
from light (compared to gauge KK masses) non-SM fermions 
that are present in models with custodial protection for the $Zb{\bar b}$ coupling 
\cite{Carena:2007tn,W'}.  Unless otherwise specified, the LHC center of mass 
energy $\sqrt{s}= 14$~TeV is assumed and the discovery reach is  
set by a 5$\sigma$ significance, in this section.

\underline{The KK gluon:} The reach for the lightest KK gluon $g^1$ 
was examined in Refs.~\cite{Agashe:2006hk,Lillie:2007yh}.  Here, the  
production is from light quark initial states, which is suppressed as discussed 
before, and the dominant decay channel is into top quarks.  The width of 
$g^1$ is roughly given by $m_{KK}/6$.  Apart from the requisite kinematic cuts, 
one could use the highly boosted top polarization as a handle on the signal.  This is 
due to the preferential coupling of $g^1$ to one of the top chiralities, set by  
the different localization of the doublet and singlet states.  The conclusion reached 
by both Refs.~\cite{Agashe:2006hk,Lillie:2007yh} indicate that with \fb{100}, $g^1$ 
up to masses of 3-4~TeV can be discovered at the LHC.  Hence, at least from the point of 
view of non-flavor constraints \cite{Carena:2007ua}, the LHC can 
probe a phenomenologically interesting regime 
for the mass of $g^1$.

\underline{The KK graviton:} The graviton KK tower is a distinct signal of the RS background.  
Refs.~\cite{Fitzpatrick:2007qr,Agashe:2007zd} revisited the LHC prospects for the discovery 
of the lightest graviton KK mode $G^1$.  The only important initial 
states for the production of $G^1$ are the gluons, with a volume-suppressed 
coupling.  However, $G^1$ has several important decay channels.  
Ref.~\cite{Fitzpatrick:2007qr} focused on the top decay channel and concluded that for top 
reconstruction efficiencies ranging over 1-100\%, the reach for $G^1$ can be 1.5-2~TeV, for 
\fb{100}.  Here, the event reconstruction is more challenging, due to the collimated  
decay products of the top, given its high boost ($E_t \sim 1$~TeV).   

Ref.~\cite{Agashe:2007zd} 
considered the process  $gg \to G^1\to Z_L Z_L \to 4 \ell$, with $\ell=e,\mu$, and 
found that with \fb{300}, $G^1$ up to a mass of about 2~TeV can 
be discovered at the LHC.  The 4-lepton final state provides for a clean signal, 
but suffers from a small branching fraction.  One may improve the signal by considering the 
hadronic final states of one of the $Z$'s.  However, the high boost of 
the $Z$ causes its dijet decay products to appear as a single jet, making $Z+j$ 
the relevant SM background, which would overwhelm the signal.  Here we note that 
the RS model predicts the mass of   $G^1$ to be $3.83/2.45\simeq 1.56$ times larger than the mass 
of the corresponding gauge KK state.  Given our previous discussion of the precision bounds, 
we then expect the mass of $G^1$ to be above 4~TeV, making its discovery a difficult challenge 
at the LHC.

\underline{The electroweak sector KK modes:}  Here, we assume that the bulk is endowed 
with a gauged $SU(2)_L\times SU(2)_R\times U(1)_X$ symmetry.  Given this gauge structure, 
at the first KK level, there are 3 neutral and 4 charged states, collectively denoted by $Z'$ and  
$W'$, respectively.

Ref.~\cite{Agashe:2007ki} considered the reach for the $Z'$.  The main decay channels here are 
into $t {\bar t}$, $W_L W_L$, and $Z_L H$.  Due to the near degeneracy of the KK gluon and $Z'$ 
masses, the top decay channel is dominated by the KK gluon ``background."  This work concluded that in  
the $Z^\prime \to W^+_L W^-_L\to \ell^+ \ell^- E_T\!\!\!\!/$ channel, the reach for the $Z'$ 
at the LHC is about 2~TeV, with \fb{100}.  The use of the $jj$ final state for on the $W$'s 
requires considering the large $W j$ SM background, since, as before, the dijet final sate will 
appear as a single jet because of the boost of the parent particle.  Ref.~\cite{W'} considered the reach for 
the $W'$, where one might expect an improved reach due to {\it (1)} the more effective reconstruction 
of the $W Z$ decay channel using clean purely leptonic final states , as there is only one neutrino 
in this case and {\it (2)} the $W'$ decay into $t {\bar b}$ is free from KK gluon contamination (unlike 
the case of the $Z'\to t {\bar t}$).  However, as it turns out, these advantages are offset by 
the smallness of the $Z$ leptonic branching ratio and 
the challenge of distinguishing a highly boosted $t$ from a $b$ (making  
$t {\bar t}$ a relevant reducible background to the $t {\bar b}$ signal).  
Hence, by and large, Ref.~\cite{W'} finds 
a reach for $W'$ similar to that for $Z'$ at the LHC.  

Many of the above conclusions about the reach of the LHC for new resonances can be 
improved by having better control over the reducible backgrounds associated with 
the decays of highly boosted heavy SM states.  We will not enter into a discussion of 
how more sophisticated analysis techniques may be employed to deal with such effects here.  
However, we mention that some recent works on top-jets and the 
substructure of high-$p_T$ jets can be found in 
Ref.~\cite{topjet} and   Ref.~\cite{Almeida:2008yp}, respectively.  Another possibility to improve 
the identification of merged dijets can be from utilization of the electromagnetic (EM) 
calorimeter, with its finer segmentation, to look for the EM cores of the jets \cite{FP}.

\section{Models without an Elementary Higgs}

Here, we will briefly summarize some key features of warped models that do not employ a
Higgs field to achieve EWSB.  We consider two classes: (1) warped Higgsless 
models  and (2) fermion condensation models. 

\underline{Warped Higgsless models:} These models \cite{Csaki:2003zu} (first introduced 
in the context of flat backgrounds \cite{Csaki:2003dt}) achieve EWSB through a set of boundary 
conditions which basically allow one to remove certain zero modes from the low energy 
spectrum.  Here, in the absence of a Higgs field, unitarity in longitudinal 
gauge boson scattering is restored via the exchange of KK modes.  However, 
this introduces an inherent tension into such a setup: achieving a reliably calculable 
model up to energies $E\gg 1$~TeV requires the KK modes to have 
$m_{KK} \lsim 1$~TeV, yet compliance with precision EW data typically demand 
$m_{KK} \gsim 1$~TeV \cite{Nomura:2003du,Barbieri:2003pr,Davoudiasl:2003me}.  
Also, these models do not provide a natural setting for obtaining a realistic 
4D flavor structure \cite{Csaki:2003sh}.

\underline{Warped quark condensation:} Given the enhanced coupling 
of KK gluons to IR-localized fermions, it is interesting to consider whether this 
coupling could lead to quark condensation and EWSB.  In particular, for a sufficiently 
IR-localized quarks $\psi_{L,R}$, one ends up with a dimension-6 operator
\[
\frac{- g_1 g_2}{M_{KK}^2}({\bar \psi_{1L}}\gamma_\mu T^a \psi_{1L})({\bar \psi
_{2R}}\gamma^\mu T^a \psi_{2R}) = \]
\beq
\frac{g_1 g_2}{M_{KK}^2}({\bar \psi_{1L}}\psi_{2R})({\bar \psi_{2R}}\psi_{1L}) +
 \ord{1/N_c},
\label{dim6}
\eeq
where $N_c=3$ is the number of colors.  The structure of the leading operator results in  
the observed pattern of EWSB upon the condensation of ${\bar \psi_{1L}}\psi_{2R}$, 
through the exchange of KK gluons.  A dynamical mass is also generated for the condensing 
quarks.  In Ref.~\cite{Burdman:2007sx}, the condensing fermions were assumed to belong to 
a fourth generation.  Here, the fourth generation fermions have masses of several hundred GeV, 
and the KK modes have $m_{KK}\gsim 1$~TeV.  
The composite Higgs is predicted to be very massive, with $m_H \gsim 700$~GeV.

Ref.~\cite{Bai:2008gm} considered a situation where  the condensing quarks 
are the top quarks of the SM.  In this model, an extra 5D quark with the quantum numbers 
of the right-handed top quark is needed to obtain realistic masses for the top quark.  Here, 
assuming that various calculable and non-calculable contributions to the radion potential 
are governed by the weak-scale, one finds that top-condensation sets the scale of 
KK masses at around 30~TeV.  The composite Higgs comes out to have a mass $m_H\sim 500$~GeV.  
The signatures of this setup are a color-charged quark 
with a mass of a few TeV and a very weakly coupled radion $\phi$ , 
with inverse coupling scale of order 100~TeV and 
a light mass of $m_\phi \lsim 4$~GeV.  Given that the radion can mediate FCNC 
processes among localized zero modes of 5D fermions \cite{Azatov:2008vm}, 
Ref.~\cite{Davoudiasl:2009xz} considered possible constraints and signals 
associated with the decay process $B \to X_s \phi$, for $m_\phi < m_B$, 
in the warped top-condensation model of Ref.~\cite{Bai:2008gm}.
 
\section{Warped Dark Matter}

Here, we survey a few proposals for a dark matter candidate in the context 
of warped 5D models. 

\underline{SO(10) GUT models:} In Ref.~\cite{Agashe:2004ci}, in order to suppress  
large violations of baryon number $B$, it was assumed to be a gauged 5D quantum 
number within a warped GUT model in which quarks and lepton come from different split 
multiplets.  This leads to a good 
$Z_3$ symmetry from a combination of $SU(3)_c$ and $B$.  Whereas the SM 
fields are not charged under $Z_3$, exotic KK particles (without zero modes) are.  
The lightest $Z_3$-charged particle (LZP), with the quantum numbers 
of a right-handed neutrino, is then stable   
in this setup and can be a good dark matter candidate.

\underline{Warped KK-parity models:}  If an extra dimension 
has symmetric boundaries, there is a discrete remnant of the original 
translation invariance that survives, referred to as KK-parity.  However, 
obviously, the warped RS background does not allow for this feature.  Hence, one 
could imagine manufacturing a symmetric warped background by gluing 
two identical copies of the RS geometry at either the UV or the IR boundary.  This 
was considered in Ref.~\cite{Agashe:2007jb}, where the first option was referred to as 
IR-UV-IR and the second one was referred to as UV-IR-UV.  However, generally speaking, 
the IR-UV-IR setup has problems related to EW constraints and flavor, whereas the UV-IR-UV 
case suffers from gravitational (radion ghost) instabilities.  

\underline{Minimal models:} An example of such a model is discussed 
in Ref.~\cite{Ponton:2008zv}, where it was shown that a scalar odd under 
a $Z_2$ symmetry can be a good  TeV-scale dark matter.

\section{Truncated RS Models} 

It was suggested, in Ref.~\cite{LRS}, that some of the constraints 
on the RS model, rooted in the strong couplings of the KK modes to IR-brane-localized  
degrees of freedom, could be loosened if one addresses 
hierarchies less severe than that between $M_P$ and the 
weak scale.  That is, 
by reducing $\sqrt{k r_c \pi}$, the aforementioned KK couplings can become less 
strong, and, for example, tree-level contributions to the oblique $T$ parameter 
can become much smaller \cite{LRS}.  
It was noted that one could still use the RS background 
to construct natural models of flavor, irrespective of the Planck-weak hierarchy.  
In this case, one may still want to keep the 
UV scale at or above $\sim 1000$~TeV (as a safe flavor scale); in Ref.~\cite{LRS} 
such truncated models were dubbed ``Little Randall-Sundrum" (LRS) models.  
Hence if the IR-brane (KK physics) is characterized by scales of 
order 1~TeV we would have LRS models with $k r_c \pi \gsim 6$.  
However, the work of Ref.~\cite{LRSepsK} has pointed out that the constraints 
from the $\epsilon_K$ parameter require a somewhat larger UV scale, corresponding to 
$k r_c \pi \gsim 7$.

An interesting aspect of volume-truncation in LRS models, pointed out in Ref.\cite{LRS}, 
is the significant sensitivity of some KK properties to the truncation parameter 
\beq
y = \frac{(k r_c)_{RS}}{(k r_c)_{LRS}}.
\label{y}
\eeq
This can affect the experimental prospects for ``Little" KK discovery at the LHC.  
To see this, let us consider the $Z'$ production and decay.  In the narrow width approximation, 
the signal $S$ for $f_1 {\bar f_1} \to Z' \to  f_2 {\bar f_2}$, is proportional to the cross section 
\beq
\sigma(f_1,f_2) \propto \Gamma_{Z'}^{f_1}\; {\rm Br}_{f_2}, 
\label{sig}
\eeq
where $\Gamma_{Z'}^{f_1}$ 
is the partial width of $Z'$ into $f_1 {\bar f_1}$, and ${\rm Br}_{f_2}$ 
is the $Z'$ branching ratio into $f_2 {\bar f_2}$.  The dominant decay modes, which control the total 
width $\Gamma_{Z'}$ of the $Z'$, are IR-localized, with couplings that are enhanced 
by a factor of order $\sqrt{k r_c \pi}$.  These dominant decay modes are also difficult 
to reconstruct and suffer from various reducible backgrounds, whereas the {\it clean} 
dilepton ($\ell^+ \ell^-$, $\ell=e,\mu$) modes have small branching 
ratios ${\rm Br}_\ell$, given their $1/\sqrt{k r_c \pi}$ suppressed couplings to KK modes, 
due to UV-localization (similar to light quarks $q$).

\begin{figure}[htb]
\includegraphics[width=7.4cm]{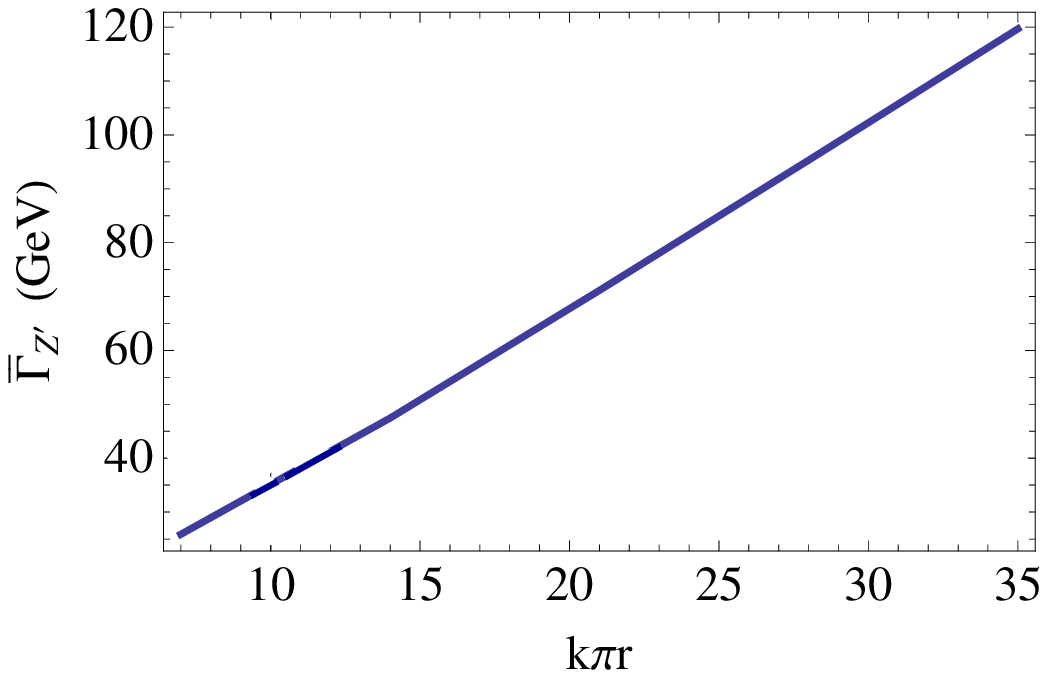}
\includegraphics[width=7.4cm]{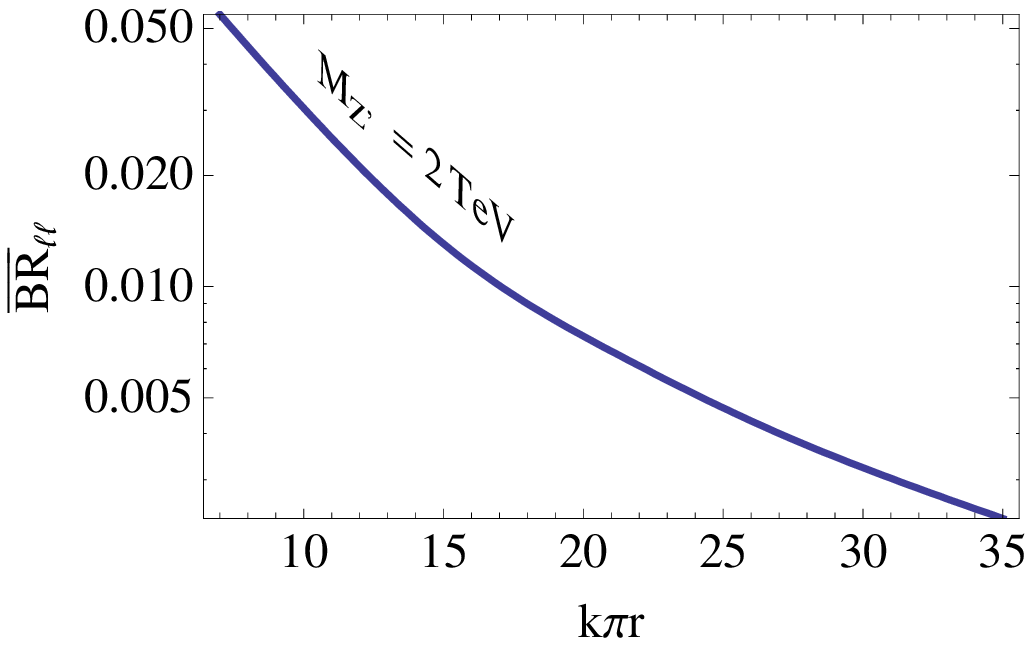}
\caption{The width of a 2-TeV $Z'$ (upper panel) and branching 
ratio into $e$ or $\mu$ pairs (lower panel), 
averaged over 3 neutral states, as a function of $kr_c \pi$; from 
Ref.~\cite{LRSatLHC}.}
\label{gambr}
\end{figure} 
\begin{figure}[htb]
\includegraphics[width=7.4cm]{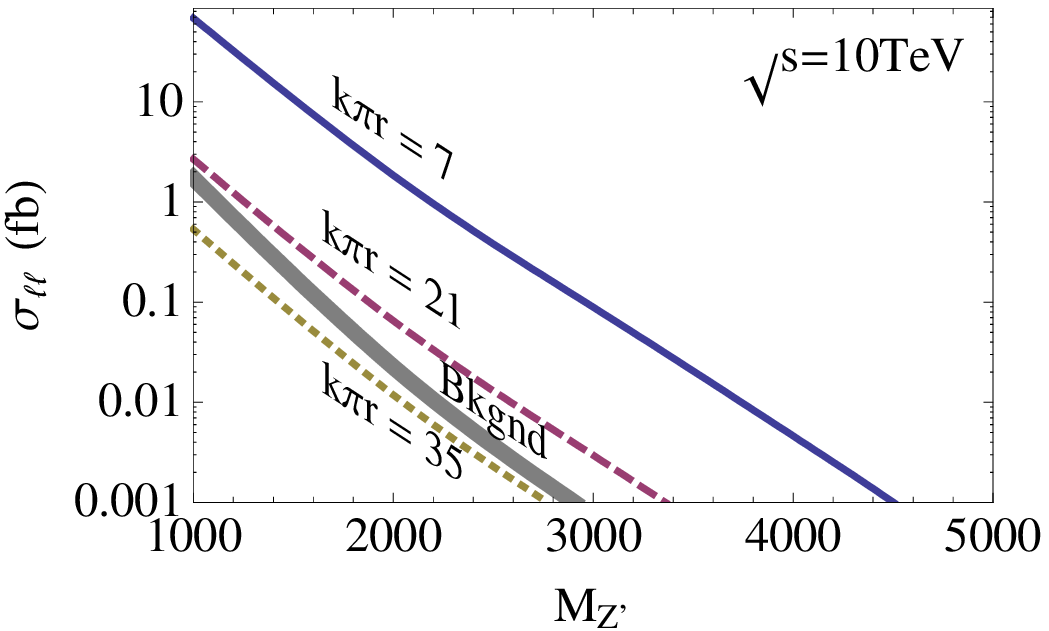}
\includegraphics[width=7.4cm]{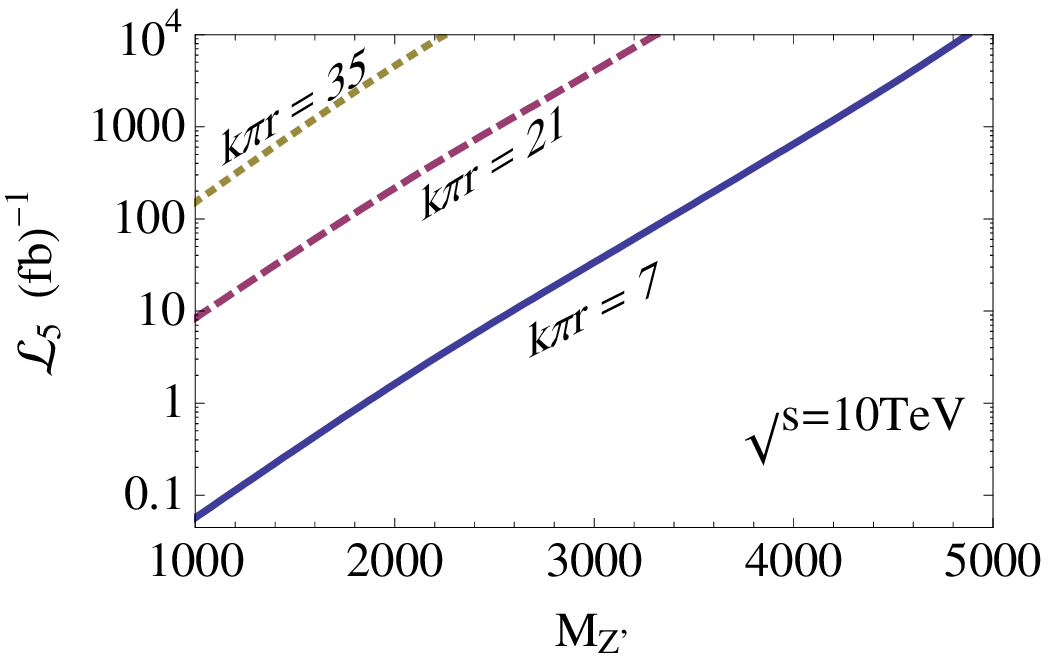}
\caption{The cross section for 
$pp\to Z'\to \ell^+\ell^-$ ($\ell=e$ or $\mu$, not both) and the SM background (upper panel), after
cuts \cite{LRSatLHC}, and the required integrated 
luminosity for a 5$\sigma$ signal with at least 3 events 
(lower panel), as a function of $M_{Z'}$.  The LHC center 
of mass energy $\sqrt{s}=10$~TeV is assumed; from 
Ref.~\cite{LRSatLHC}.}
\label{10TeV}
\end{figure} 
In warped models of flavor, we then have $\Gamma_{Z'}^f \sim y$, for $f=q, \ell$, 
$\Gamma_{Z'} \sim 1/y$, and hence ${\rm Br}_\ell \sim y^2$.  Consequently, 
we expect that 
\beq
S\propto \sigma(q,\ell) \sim y^3
\label{S}
\eeq
 and the background under the resonance peak,
$
B \propto \Gamma_{Z'}\sim 1/y
$.  We hence find  
\beq
S/B \sim y^4\,.
\label{S/B}
\eeq
\begin{figure}[htb]
\includegraphics[width=7.4cm]{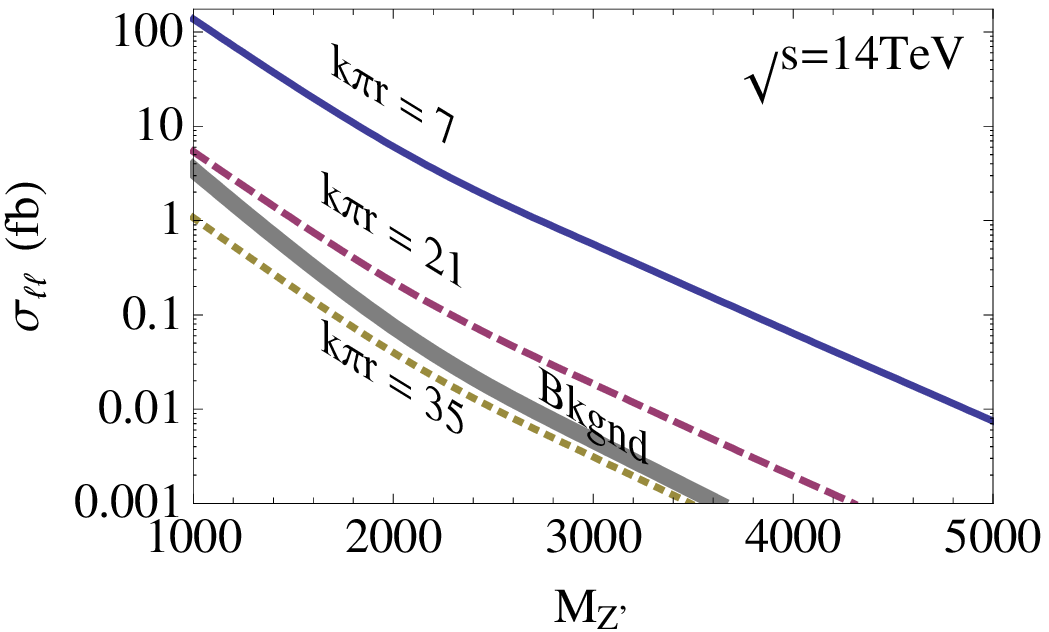}
\includegraphics[width=7.4cm]{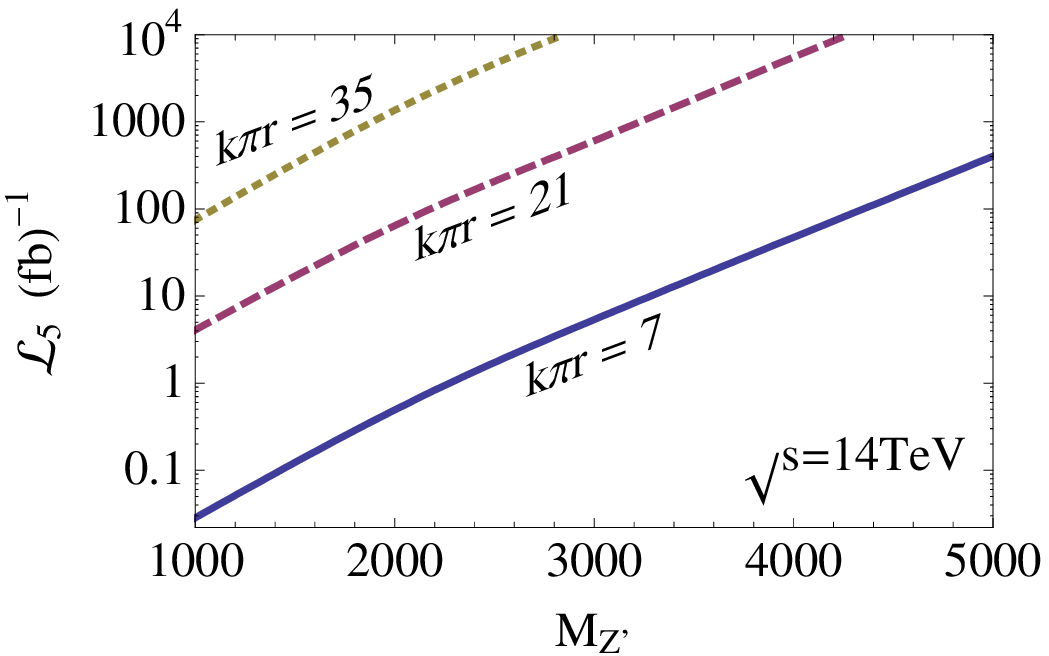}
\caption{Same as Fig.~\ref{10TeV}, but for $\sqrt{s}=14$~TeV; from 
Ref.~\cite{LRSatLHC}.}
\label{14TeV}
\end{figure} 
The remarkable enhancement in the signal size and quality for the clean 
dilepton decay modes of the $Z'$ at the LHC, as suggested by the above simple analysis, 
points to an interesting possibility.  That is, given the considerable sensitivity of 
important $Z'$ branching ratios on the truncation parameter $y$, one may use 
TeV-scale data to probe the UV scale of the theory!  Since the discovery of a 
KK-like resonance at the LHC does not necessarily point to a resolution
of the Planck-weak hierarchy, such experimental handles on the UV-brane scale 
in the LRS models are quite valuable in deciphering the microscopic nature of 
the underlying theory.

A more quantitative examination of the effect of truncation on the LHC 
signals of warping is presented in Ref.~\cite{LRSatLHC} and we will focus 
on the $Z'$ signals from this work.  All the subsequent plots shown here are from 
Ref.~\cite{LRSatLHC}.  The width of a 2-TeV $Z'$ and its 
branching ratio into $\ell^+ \ell^-$ ($\ell=e$ or $\mu$, not both) as a function of $k r_c \pi$, 
averaged over 3 states, are given in Fig.~\ref{gambr}.  These results confirm the expected behavior 
under truncation.

In Fig.~\ref{10TeV}, as a function of $M_{Z'}$, the cross section (upper panel) 
for $pp \to Z' \to \ell^+\ell^-$ at the LHC with $\sqrt{s}=10$~TeV, after suitable cuts \cite{LRSatLHC}, 
and the required integrated luminosity for a $5\sigma$ signal 
with at least 3 events (lower panel), for $kr_c \pi =7,21, 35$, are presented.  These results suggest that,
for $k r_c \pi \approx 7$, a 2-TeV 
Little $Z'$ can be discovered in this clean decay channel, at $\sqrt{s}=10$~TeV, with only \fb{1}.  This 
shows that phenomenologically interesting values of the UV brane scale, 
corresponding to $M_5\sim 10^4$ TeV, 
can be probed with {\it early} LHC data.  The same quantities are shown for $\sqrt{s}=14$~TeV, as 
may be expected at later stages of the LHC running, in Fig.~\ref{14TeV}.  
Here, we can deduce that a 3-TeV Little $Z'$ 
($kr_c \pi \approx 7$) can be discovered in the clean dilepton channel with only \fb{4} 
of integrated luminosity.  This should be compared with the equivalent discovery reach requiring 
\fb{300}, in {\it any channel}, if $M_5\sim M_P$, as in the original RS model.  Hence, we see that 
truncation has a significant effect on the reach for warped KK modes and TeV information on such 
states can shed light on the UV scale of the 5D model.

\section{Conclusions}

Warped 5D models can provide a predictive framework to explain the hierarchy and 
flavor puzzles of the SM, and some of their variants can 
also provide good dark matter candidates.  In these models the gauge and matter fermion 
content of the SM is placed in all 5D.  Avoiding fine-tuning of parameters requires 
the introduction of extended 5D gauge symmetries, to protect precision EW observables from 
large corrections.  With such protective symmetries, the masses of the KK modes can be in 
2-3 TeV range, however flavor generally requires additional structure to avoid pushing 
KK modes to above $\sim 10$~TeV.  The warped models that provide realistic 4D flavor 
pose a challenge to LHC experiments, as they are characterized by 
suppressed couplings between the KK modes and light SM fields (important for production and 
detection).  

However, if one relaxes the requirement of addressing  the Planck-weak hierarchy, 
one could still obtain natural models of flavor, with fundamental 5D scales at or above 
$\sim 10^4$~TeV.  These truncated ``Little Randall-Sundrum" models address the hierarchy 
between the high ``flavor" scale and the weak scale 
(where the KK modes appear), but have much improved LHC discovery 
prospects.  The sensitivity of the KK physics to truncation 
provides an opportunity for probing reasonable values of the 5D scale of 
the underlying geometry, starting with early LHC data.  

\section*{Acknowledgments}

We would like to thank the organizers of the 2009 BSM-LHC workshop for their invitation to 
give the talk upon which this writeup is based.

\end{document}